# Web-page Indexing based on the Prioritize Ontology Terms

Sukanta Sinha[1, 4], Rana Dattagupta[2], Debajyoti Mukhopadhyay[3, 4]

[1]Tata Consultancy Services Ltd., Victoria Park Building, Salt Lake, Kolkata 700091, India
sukantasinha2003@gmail.com
[2]Computer Science Dept., Jadavpur University, Kolkata 700032, India
ranadattagupta@yahoo.com
[3]Information Technology Dept., Maharashtra Institute of Technology, Pune 411038, India
debajyoti.mukhopadhyay@gmail.com
[4]WIDiCoReL Research Lab, Green Tower, C-9/1, Golf Green, Kolkata 700095, India

**Abstract.** In this world, globalization has become a basic and most popular human trend. To globalize information, people are going to publish the documents in the internet. As a result, information volume of internet has become huge. To handle that huge volume of information, Web searcher uses search engines. The Web-page indexing mechanism of a search engine plays a big role to retrieve Web search results in a faster way from the huge volume of Web resources. Web researchers have introduced various types of Web-page indexing mechanism to retrieve Web-pages from Web-page repository. In this paper, we have illustrated a new approach of design and development of Web-page indexing. The proposed Web-page indexing mechanism has applied on domain specific Web-pages and we have identified the Web-page domain based on an Ontology. In our approach, first we prioritize the Ontology terms that exist in the Web-page content then apply our own indexing mechanism to index that Web-page. The main advantage of storing an index is to optimize the speed and performance while finding relevant documents from the domain specific search engine storage area for a user given search query.

**Keywords:** Domain Specific Search, Ontology, Ontology Based Search, Relevance Value, Search engine, Web-page Indexing

## 1 Introduction

In recent years, the growth of the World Wide Web (WWW) has been rising at an alarming rate and contains a huge amount of multi-domain data [1]. As a result, there is an explosion in information and web searcher uses search engines to handle that information. There are various parameters used by the search engines to produce better search engine performance, Web-page indexing is one of them. Nowadays, Web researchers have already introduced some efficient Web-page indexing mechanism like Back-of-the-book-style Web-page indexes formally called "Web site A-Z index-es", "Human-produced Web-page index", "Meta search Web-page indexing", "Cache based Web-page indexing", etc. [2].

In our approach, we have introduced a new mechanism for Web-page indexing. This is fully domain specific Ontological approach, where each Ontology term is treated as a base index. Ontology index number assigned based on their weight value [3-4]. In our proposed mechanism, first we retrieve dominating and sub-dominating Ontology terms for a considered Web-page from the domain specific Web-page repository, then apply primary and secondary attachment rule according to our proposed mechanism.

The paper is organized in the following way. In section 2, we have discussed the related work on Web-page indexing. The proposed architecture for domain-specific Web-page indexing is given in section 3. All the component of our architecture is also discussed in the same section. Experimental analyses and conclusion of our paper is given in section 4 and 5 respectively.

**Definition 1.1: Dominating Ontology Term-** Ontology term which holds maximum Ontology term relevance value in the considered Web-page.

**Definition 1.2: Sub-dominating Ontology Terms-** Ontology terms which hold successive maximum Ontology term relevance values other than dominating Ontology term in the considered Web-page.

**Rule 1.1: Primary Attachment (P1, P2 …) –** All the dominating Ontology terms for all Web-pages are indexed with the primary attachment of their respective Ontology term.

**Rule 1.2: Secondary Attachment (S1, S2 …) -** All the sub-dominating Ontology terms for all Web-pages are indexed with the secondary attachment of their respective Ontology term.

## 2      Related Works

The main advantage of storing an index is to optimize the speed and performance while finding relevant documents from the search engine storage area for a user given search criteria. In this section, we are going to discuss the existing Web-page indexing mechanism and their drawbacks.

**Definition 2.1: Ontology –**It is a set of domain related key information, which is kept in an organized way based on their importance.

**Definition 2.2: Relevance Value –**It is a numeric value for each Web-page, which is generated on the basis of the term Weight value, term Synonyms, number of occurrences of Ontology terms which are existing in that Web-page.

**Definition 2.3: Seed URL –**It is a set of base URL from where the crawler starts to crawl down the Web pages from the Internet.

**Definition 2.4: Weight Table –** This table has two columns, first column denotes Ontology terms and second column denotes weight value of that Ontology term. Ontology term weight value lies between '0' and '1'.

**Definition 2.5: Syntable -** This table has two columns, first column denotes Ontology terms and second column denotes synonym of that ontology term. For a particular ontology term, if more than one synonym exists, those are kept using comma (,) separator.

**Definition 2.6: Relevance Limit –**It is a predefined static relevance cut-off value to recognize whether a Web-page is domain specific or not.

**Definition 2.7: Term Relevance Value –** It is a numeric value for each Ontology term, which is generated on the basis of the term Weight value, term Synonyms, number of occurrences of that Ontology term in the considered Web-page.

### 2.1    Back-of-the-book-style

Back-of-the-book-style Web-page indexes formally called "Web site A-Z indexes". Web site A-Z indexes have several advantages. But search engines language is full of homographs and synonyms and not all the references found will be relevant. For example, a computer-produced index of the 9/11 report showed many references to George Bush, but did not distinguish between "George H. W. Bush" and "George W. Bush" [5].

### 2.2    Human-produced Web-page Index

Human-produced index has someone check each and every part of the text to find everything relevant to the search term, while a search engine leaves the responsibility for finding the information with the enquirer. It will increase miss and hit ratio. This approach is not suitable for the huge volume of Web data [6].

### 2.3    Meta Search Web-page Indexing

Metadata Web indexing involves assigning keywords or phrases to Web-pages or websites within a meta-tag field, so that the Web-page or website can be retrieved by a search engine that is customized to search the keywords field. This may be involved using keywords restricted to a controlled vocabulary list [7].

### 2.4    Cache based Web-page Indexing

Frequently used search query produces search result quickly because the result information stored into cache memory. On the other hand while an irregular search string encountered, the search engine cannot produce faster search result due to information not available in the cache memory. Irregular search strings always come because of the huge volume of internet information and user [8-9].

## 3 Proposed Approach

In our approach, we have proposed a new mechanism for indexing domain specific Web-pages. Before going forward with the new indexing mechanism, we need to make sure all the inputs are available in our hands. Those inputs are domain specific Web-page repository, set of Ontology terms, Weight table and Syntable [10]. One of our earlier work, we have created the domain specific Web-page repository [11-12]. We have used that repository as an input of our proposed approach.

### 3.1 Extraction of Dominating and Sub-Dominating Ontology Terms

In this section, we will discuss how to extract dominating and sub-dominating Ontology terms. We will illustrate this by using one example (refer Fig. 1).

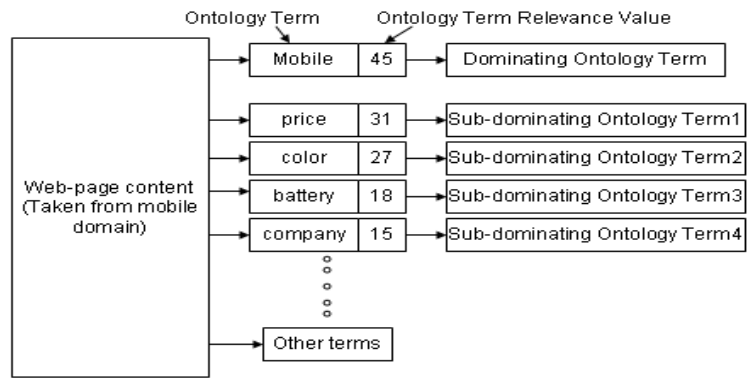

**Fig. 1.** Example of Extracting Dominating and Sub-dominating Ontology Terms

Consider a 'Mobile' domain Web-page. First extract the Web-page content then apply definition 1.1 and 1.2. We have found that Ontology term 'Mobile' holds term relevance value 45, which is maximum and according to our definition 1.1 Ontology term 'Mobile' becomes dominating Ontology term. Ontology term 'price', 'color', 'battery' and 'company' holds term relevance value 31, 27, 18 and 15 respectively, which are greater than all other Ontology terms excluding 'Mobile' Ontology term. Now according to our definition 1.2, Ontology term 'price', 'color', 'battery' and 'company' become sub-dominating Ontology term 1, sub-dominating Ontology term 2, sub-dominating Ontology term 3 and sub-dominating Ontology term 4 respectively. If number of sub-dominating Ontology term increased then secondary attachments also increases proportionally to store them (refer Rule 1.2), which increases indexing memory size. For that reason, we have used four sub-dominating Ontology terms as a threshold value. Some rare cases, we found multiple Ontology term holds same term relevance value that time we will prioritize dominating and sub-dominating Ontology terms according to their lower term weight value, i.e., consider the higher value of the number of occurrences of that Ontology term in the considered Web-page content.

## 3.2 Proposed Algorithm of Web-page indexing

Proposed algorithm briefly describes the mechanism of Web-page indexing based on the prioritized Ontology terms for a set of domain specific Web-pages.

```
Input   : Domain specific Web-pages
Output  : Indexed all the Web-pages
```

1. Select a Web-page (P) from domain specific Web-page repository
2. Extract Dominating Ontology Term (D)
3. Extract Sub-Dominating Ontology Terms (SDi where 0<i≤4 and i is an integer)
4. Add Web-page identifier (P_ID) of P with Primary attachment of D
5. Add Web-page identifier (P_ID) of P with Secondary attachment of SDi where 0<i≤4 and i is an integer
6. Repeat step 1-5 until all the Web-pages get indexed
7. End

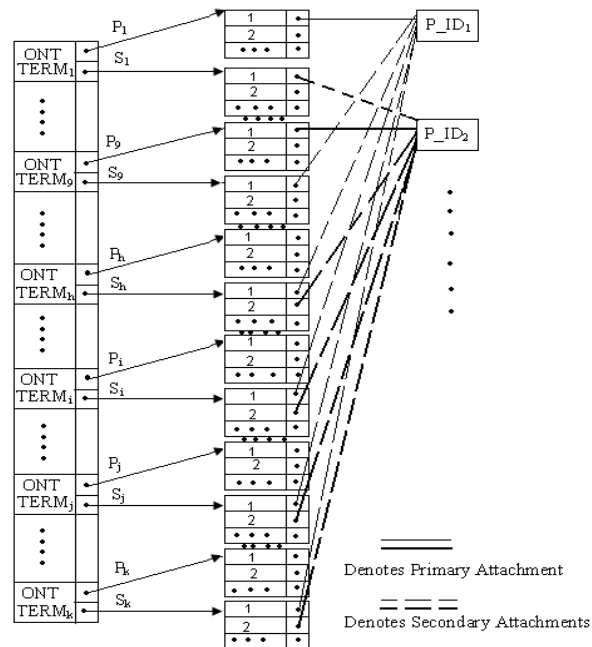

**Fig. 2.** Web-page structures after applying our indexing mechanism

A pictorial diagram of Web-page structures after applying our indexing mechanism is shown in fig. 2. Each Ontology term maintains two tables. One table used for storing primary attachments and other one used for storing secondary attachments (refer Rule 1.1 and 1.2). In fig. 2, ($P_1$,…, $P_9$,…, $P_h$, …, $P_k$) and ($S_1$,…, $S_9$,…, $S_h$, … , $S_k$) all are pointing primary and secondary attachment table of their corresponding Ontology terms respectively. Each Web-page has only one primary attachment and four secondary attachments. In the fig. 2, (P_ID$_1$, P_ID$_2$,....) representing Web-page identifier of each considered domain specific Web-pages. Solid lines are denoting primary attachment, which pointing primary attachment of dominating Ontology term. Dotted lines are denoting secondary attachments, which pointing secondary attachment of sub-dominating Ontology terms.

### 3.3 User Interface

In our proposed search engine, we have facilitated Web searchers to customize their search result by selecting all the inputs. We have used drop-down lists for selecting dominating and sub-dominating Ontology terms. Web-searcher can produce optimistic search results from our proposed search engine without knowing the domain knowledge because all the Ontology terms are already available in the drop-down lists. After providing all the inputs, i.e., search tokens, relevance range and number of search results, Web searcher need to click on "Search" button to get the search results. In fig.3 shows a part of the user interface of our prototype and '*' denotes mandatory fields. 'Number of Search Results' field restricts the Web searcher produce limited search result. For an example, say 100 search results are produced for user given search tokens and relevance rage, now user wants 20 search results that time user needs to put 20 in the 'Number of Search Results' field. Lesser time will be taken to displaying 20 result links instead of displaying 100 result links. In the user interface, the maximum relevance value and minimum relevance value are set dynamically according to the practical scenario based data or query.

**Fig. 3.** A Part of User Interface

### 3.4 Web-page Retrieval Mechanism Based on the User Input

Web-page retrieval from Web search engine resources are an important role of a Web search engine. We are retrieving a resultant Web-page list from our data store based on the user given dominating and sub-dominating Ontology terms, relevance range, etc. Most of the cases in the existing search engines follow to parse the search string and then retrieve the Web-pages based on those parsed tokens. According to our prototype, we are giving a flexibility to the user does not use the search string, directly select the search tokens from the drop down lists (refer Fig. 3). As a result, it reduces the search string parsing time and miss hit ratio due to user's inadequate domain knowledge. As discussed in section 3.3, at a time user can select only one dominating and four sub-dominating Ontology terms. Our prototype uses below formula to produce a resultant Web-page list based on the user given relevance range.

(50% of 'x' from the primary attachment list of dominating Ontology term +
20% of 'x' from secondary attachment list of first sub-dominating Ontology term +
15% of 'x' from secondary attachment list of second sub-dominating Ontology term +
10% of 'x' from secondary attachment list of third sub-dominating Ontology term +
5% of 'x' from secondary attachment list of fourth sub-dominating Ontology term),
where 'x' denotes 'Number of Search Results' in the user interface (refer Fig. 3).

## 4       Experimental Analyses

In this section, we have given some experimental study as well as discussed how to set up our system. Section 4.1 explains our experimental procedure, section 4.2 depicts our prototype time complexity to produce resultant Web-page list and section 4.3 shows the experimental results of our system.

### 4.1    Experiment Procedure

Performance of our system depends on various parameters, and those parameters need to be setup before running our system. The considered parameters are domain relevance limit, weight value assignment, Ontology terms, domain specific Web-page repository, etc. These parameters are assigned by tuning our system through experiments. We have created domain specific Web-page repository by taking 50 seed URLs is an input of our domain specific Web search crawler.

### 4.2    Time Complexity to Produce Resultant Web-page List

We have considered 'k' number of Ontology terms. We have kept them in a sorted order according to their weight value. While finding user given dominating Ontology term primary attachment link, our prototype required maximum $O(\log_2 k)$ time using binary search mechanism (refer Fig. 2). On the other hand while finding other four user given sub-dominating Ontology term secondary attachment links, our prototype required $4O(\log_2 k)$ times. In the second level, our prototype reached from primary and secondary attachment to the Web-pages just spending constant time because there is no iteration required. Finally, our prototype time complexity becomes $[5O(\log_2 k) +5c] \approx O(\log_2 k)$ to the retrieve resultant Web-page list, where 'c' is a constant time required to reach the primary and secondary attachment to Web-pages.

### 4.3 Experimental Result

It is very difficult to compare our search results with the existing search engines. Most of the cases, existing search engines do not hold domain specific concepts. It is very important that while comparing two systems both are on the same page, i.e., contains same resources, environment, system platforms, search query all are same. Few existing cases, where search engine gives an advanced search option to the Web searchers, but not match with our domains. Anyhow we have produced few data to measure our proposed prototype performance. To produce the experimental results, we have compared the two systems (before and after applying Web-page indexing mechanism) performances. In table 1, we have given a performance report of our system.

**Table 1** Performance Report of Our System

| Number of Search Results | Time Taken (in Seconds) | | Total Number of Web-pages in the Repository |
|---|---|---|---|
| | Before applying Web-page indexing | After applying Web-page indexing | |
| 10 | 0.530973451 | 0.392156863 | 5000 |
| 20 | 1.085972851 | 0.860215054 | 5000 |
| 30 | 1.753246753 | 1.409921671 | 5000 |
| 40 | 2.394014963 | 2.008368201 | 5000 |
| 50 | 3.018108652 | 2.683363148 | 5000 |

To measure accuracy, we have applied our set of search query multiple times by taking 'Number of Search Results' (refer Fig. 3) field values 10, 20, 30, 40 and 50. In table 2, we have shown our system accuracy measurement.

**Table 2** Accuracy of Our System

| Number of Search Results | Avg. No. of Relevant Results | Avg. No. of Non-Relevant Results | Total Number of Web-pages in the Repository |
|---|---|---|---|
| 10 | 8.7 | 1.3 | 5000 |
| 20 | 17.2 | 2.8 | 5000 |
| 30 | 26.4 | 3.6 | 5000 |
| 40 | 34.6 | 5.4 | 5000 |
| 50 | 43.6 | 6.4 | 5000 |

## 5 Conclusions

In this paper, we have proposed a prototype of a domain specific Web search engine. This prototype has used one dominating and four sub-dominating Ontology terms to produce Web search results. All the Web-pages are indexed according to their dominating and sub-dominating Ontology terms. According to our experimental results, Web-page indexing mechanism produced faster result for the user selected dominating and sub-dominating Ontology terms. According to our prototype, we are giving a flexibility to the user does not use the search string, directly select the search tokens from the drop down lists. As a result, it reduces the search string parsing time and miss hit ratio due to user's inadequate domain knowledge. This prototype is highly scalable. Suppose, we need to increase the supporting domains for our prototype, then we need to include the new domain Ontology and other details like weight table, syntable, etc. of that Ontology. In a single domain there does not exist huge number of ontology terms. Hence, the number of indexes should be lesser than a general search engine. As a result, we can reach the web-pages quickly as well as reducing index storage cost.